# Buckyball Sandwiches

R. Mirzayev , K. Mustonen [a)], M.R.A. Monazam, A. Mittelberger, T.J. Pennycook, C. Mangler, T. Susi, J. Kotakoski and J.C. Meyer [b)]

*University of Vienna, Faculty of Physics, Boltzmanngasse 5, A-1090 Vienna, Austria*

---

Electronic mail: [a)] kimmo.mustonen@univie.ac.at, [b)] jannik.meyer@univie.ac.at

These authors contributed equally to this work

**Two-dimensional (2D) materials have considerably expanded the field of materials science in the last decade. Even more recently, various 2D materials have been assembled into vertical van der Waals heterostacks, and it has been proposed to combine them with other low-dimensional structures to create new materials with hybridized properties. Here, we demonstrate the first direct images of a suspended 0D/2D heterostructure incorporating $C_{60}$ molecules between two graphene layers in a *buckyball sandwich* structure. We find clean and ordered $C_{60}$ islands with thicknesses down to one molecule, shielded by the graphene layers from the microscope vacuum and partially protected from radiation damage during scanning transmission electron microscopy imaging. The sandwich structure serves as a 2D nanoscale reaction chamber allowing the analysis of the structure of the molecules and their dynamics at atomic resolution.**

# INTRODUCTION

Graphene (*1*), a single layer of carbon in a hexagonal lattice, is the thinnest imaginable membrane, with unique intrinsic properties. It has no bulk, and interfaces with other materials through van der Waals (vdW) forces. This has made graphene appealing as one of the main components for so-called vdW heterostructures consisting either of only two-dimensional (2D) materials (*2*) or mixtures of 2D materials with structures of other dimensionalities (*2, 3*). One of the earliest examples of mixed-dimensional hybrid structures was the *carbon peapod* (*4*) in which a linear array of (quasi-)zero-dimensional fullerenes (*5*) is confined inside the (quasi-)one-dimensional carbon nanotube. Recently, collapsed double-walled carbon nanotubes have been intercalated with $C_{60}$ to form a nanoribbon type peapod (*6, 7*).

Fullerenes have many interesting properties: they have a band gap under ambient conditions (*8*), can become metallic when doped or even exhibit superconductivity under certain conditions (*9*). Fullerenes belong to the rare group of saturable absorbers and can exhibit peculiar ferromagnetic behavior at low temperatures and when polymerized (*10-12*). They are commonly synthesized by ablating graphite, where it is thought that defects cause the carbon lattice to curve into minimum-energy spheroids (*13*). In the carbon peapods, fullerenes can be studied in a confined and well-defined space without the need of a suspension. This has made it possible to directly image their dynamics and bonding under atomic resolution electron microscopy (*14, 15*). However, on surfaces fullerenes have also been reported to self-assemble into hexagonally close-packed islands, and exhibit epitaxial ordering on graphene and even on some transition metals (*16, 17*).

Here, we present the first example of suspended *buckyball sandwiches*, where $C_{60}$ molecules are confined between two graphene monolayers. The resulting structures contain large atomically clean areas, allowing the direct study of $C_{60}$ islands through scanning transmission electron microscopy (STEM). Like carbon peapods, the buckyball sandwiches are expected to have hybridized properties that could be suited for applications ranging from nanoscale lasers to spin cubit arrays and nanoscale mechanical elements (*18, 19*). Furthermore, alkali metal co-intercalation may turn it into a high temperature superconductor (*20, 21*), similar to crystalline $C_{60}$ (*22, 23*). We concentrate on understanding the atomic-scale structure of the sandwiches and the dynamics of the contained fullerenes. We observe the diffusion of individual $C_{60}$ and find that they remain rotationally active during room temperature experiments. However, during continuous electron irradiation, some $C_{60}$ first bond to form dimers and finally fused peanut-like structures, which continue to rotate around the joint axis. This movement is hindered only for structures involving three or more molecules.

Our results show that individual molecules in a graphene sandwich can be imaged at atomic resolution. Previously, graphene encapsulation has been used to study small metal particles (*24*), colloidal nanocrystals (*25*) or other 2D materials systems (*26, 27*) through high-resolution electron microscopy. Encapsulation is known to reduce the rate of radiation damage (*26, 27*), and provide a sample support with a periodic structure that can be easily subtracted from the image. As an additional important feature, we find that the graphene sandwich provides a "clean" view onto the molecules; the sandwich has only such contamination as is typical for graphene in transmission electron microscopy (TEM) imaging, while without the sandwich no clear images of the fullerenes could be obtained. The sandwich further provides a nanoscale reaction chamber that is less constrained than the 0D fullerene cage (*28*) or the 1D "test tube" inside a carbon nanotube (*15*) to study the structure and properties of individual molecules trapped between the layers.

**RESULTS AND DISCUSSIONS**

The buckyball sandwiches were fabricated by thermally evaporating $C_{60}$ onto commercially available graphene-coated TEM grids, followed by the placement of a second graphene-coated grid on top and bringing the two films into contact by evaporating a droplet of solvent (see Methods; Supplementary Figures S1 and S2 show the general sample morphology, Raman spectrum of the sandwich, and the structure of the multilayers). The sandwiched $C_{60}$ show remarkable stability under the electron beam, similar to fullerenes encapsulated in carbon nanotubes (*14*). On the basis of high-resolution images, it is easy to distinguish between mono- and multilayer regions, with an example shown in Figure 1a. Figure 1b shows an atomically resolved close-up and Figure 1c its Fourier transform, from which the two graphene layers (with a misorientation angle $\Theta_{Gr} \approx 11°$) and the fullerene lattice can be identified. Unlike on a single layer of graphene, the sandwiches also contain fullerene monolayers, whose edges exhibit non-linear image contrast of the moiré pattern of the suspended graphene layers (Supplementary Figure S3).

Using the graphene lattice as a calibration standard, we measured the $C_{60}$ lattice spacing (the distance between the centers of the neighboring molecules) for crystalline monolayer regions. Surprisingly, this spacing (9.6±0.1 Å) is 4-5% smaller than that of 3D crystallites (*29*). This distance is also smaller than in 1D carbon peapods, whose reported values vary in the range of 9.7-10.0 Å (possibly influenced by the endohedral metal atoms present in some studies) (*14, 30, 31*).

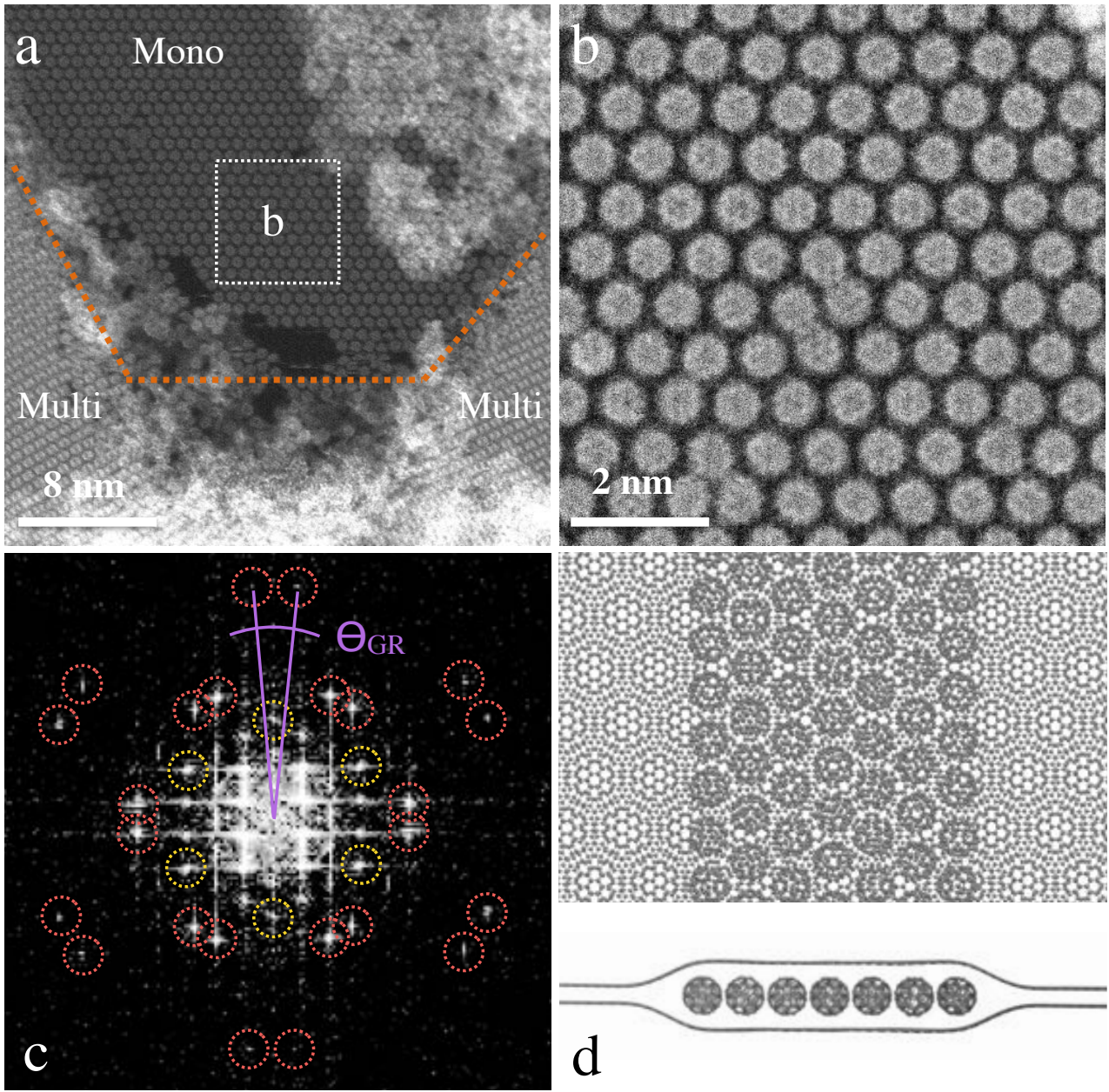

Fig. 1. Monolayer area of a buckyball sandwich. **a)** Sandwich with both mono- and multilayered $C_{60}$ regions and **b)** a higher magnification of the region labeled [b]. **c)** Fourier transform of (b), with graphene reflections marked by red dashed lines, and some of the fullerene peaks marked in yellow. **d)** Model of a monolayer $C_{60}$ sandwich (top and side view).

To understand this change in spacing, we used atomistic simulations to calculate the equilibrium lattice constants of the molecules in bulk 3D and 2D $C_{60}$ crystals, and in a 2D sandwich (see Methods). The simulated sandwich consisted of two graphene nanoribbons with an orientation mismatch of $\Theta_{Gr} \approx 11°$, with a 5×20 array of $C_{60}$ in between (Figure 1d). For 3D $C_{60}$, we found an equilibrium distance of 9.65 Å, whereas for an infinite 2D sheet, this distance was 9.62 Å. Hence, the mere dimensionality of the crystallites is unable to explain the observed monolayer spacing. However, in the simulated sandwich, the average equilibrium distance was 9.22 Å (with a standard deviation 0.15 Å due to directional anisotropy), which represents a contraction of 4.7% compared with the bulk value. Thus, despite the difference in the absolute distances, the interaction with the graphene sheets in the simulated sandwich compresses the $C_{60}$ lattice in excellent agreement with the experiment.

Individual fullerenes are found to be highly mobile at the unconstrained edges of the $C_{60}$ monolayers: some $C_{60}$ in the partially filled outermost row of Figure 2a have occupied and vacated edge sites multiple times during the acquisition of subsequent lines of the scanned image, resulting in the "interlaced" appearance of the fullerenes. For example, the fullerene marked by an arrow has disappeared and re-appeared at least 17 times during the acquisition of 100 scan-lines (~1 s) in this region of the image. However, it typically remained stationary during the ~1 ms the probe took to traverse the molecule in each line of the scan. Evidently, it is possible to quantify even single-molecule diffusion in the sandwich structure using direct STEM imaging. Regardless of their mobility inside this area, it appears the molecules remain confined in the graphene wrap and cannot completely escape.

To understand these dynamics, we also calculated the relevant diffusion barriers using the nudged elastic band method (Figure 2b; see Methods and Supplementary Videos 1 and 2): one for a $C_{60}$ diffusing along the edge of the monolayer (234 meV), and another for a $C_{60}$ diffusing from one edge to another across the gap (371 meV). These values are in reasonable agreement with those reported for fullerene diffusion on $C_{60}$ crystallites (*32, 33*). Such energies are easily available from the electron beam, but these transitions might also be thermally activated at room temperature. From the second path, we could also estimate the barrier for moving the $C_{60}$ inside the gap in the sandwich to be less than 5 meV. This explains why isolated fullerenes are never observed in the experimental images.

Different dynamics are observed in disordered $C_{60}$ regions, as shown in Figures 2c-d: in this case, voids in the $C_{60}$ arrangement can propagate inside the lattice. For example, between the two shown frames, a void propagated from location 1 to 2, and a single molecule escaped outside the field of view from position 3. Overall, however, this edge appears to be more tightly constrained by the two graphene layers, displaying no oscillations like those seen in Figure 2a.

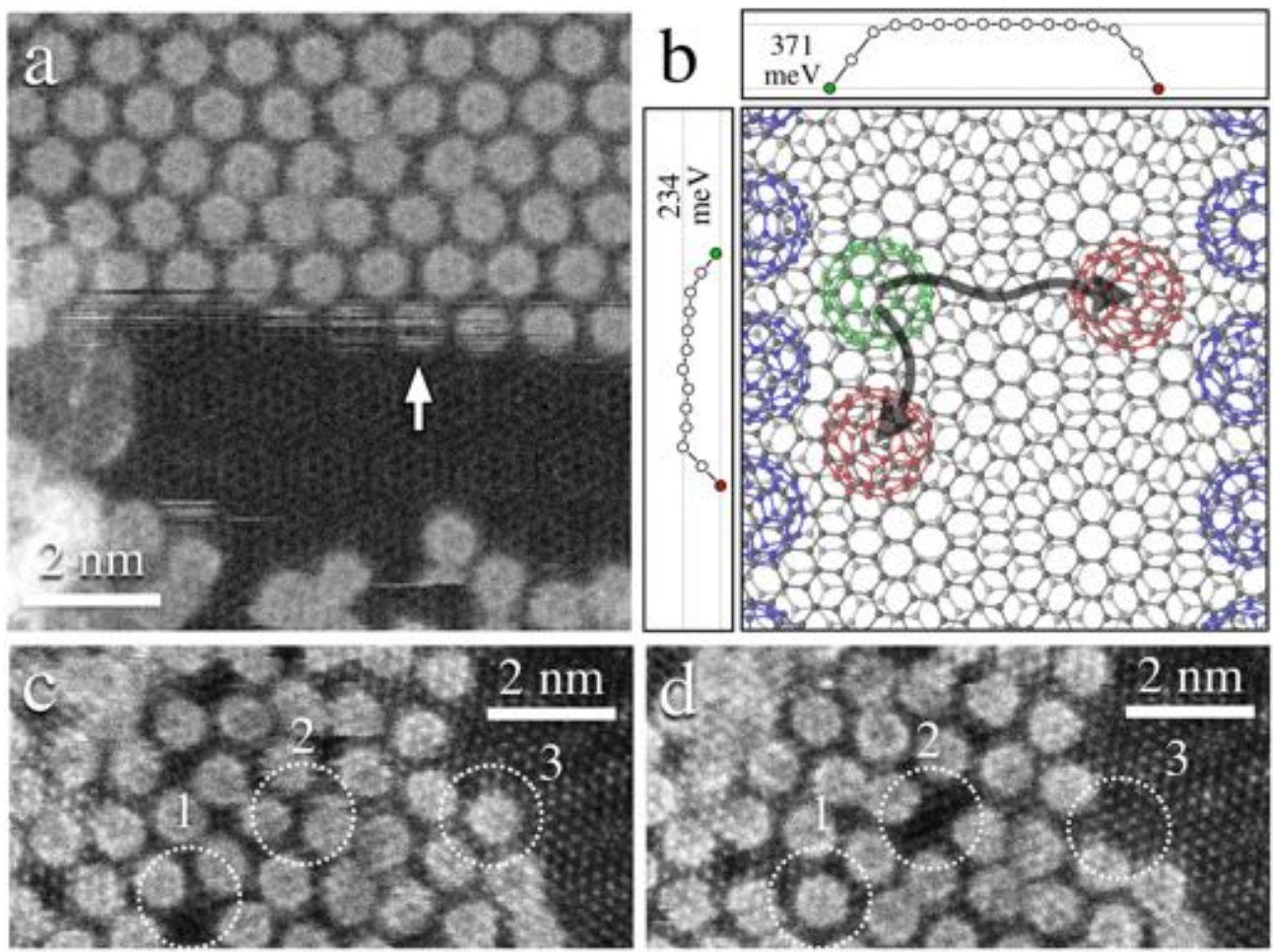

Fig 2. Dynamics of sandwiched fullerenes. **a)** An oscillating $C_{60}$ at the edge of a gap in the fullerene monolayer, over which the graphene layers are suspended. At the location indicated by the arrow, a molecule disappears and appears at least 17 times during 100 scan lines. **b)** Nudged elastic band paths and energy barriers for $C_{60}$ diffusion along an edge of fullerene monolayer (vertical) and from one edge of the gap to another (horizontal). **c-d)** A disordered monolayer showing a void propagating from location 1 to 2, and the $C_{60}$ at location 3 escaping outside the field of view between the two consecutive frames.

Next, we consider the rotation and bonding of sandwiched $C_{60}$ molecules during imaging with 60 keV electrons. At least three types of modifications are possible: knock-on damage of the $C_{60}$ (initially resulting in the formation of $C_{59}$ (*34*), with a reactive dangling bond next to the vacancy), bonding between fullerenes, or the bonding of $C_{60}$ to graphene; graphene knock-on damage is completely suppressed at this energy (*35*). Although the graphene lattice is clearly resolved in the original images of the time/electron dose sequence of Figure 3a-c (see Supplementary Material), no structure beyond spherical symmetry can be discerned for isolated fullerenes at the beginning of the sequence. Since the absorption energy of $C_{60}$ on graphene is around 0.9 eV (*36*) and different absorption geometries are separated by <0.1 eV (*16*), even the graphene sandwich should not prevent $C_{60}$ rotation under our experimental conditions. We therefore attribute the lack of internal structure of the $C_{60}$ in our images to rotation, either thermally activated or driven by the electron irradiation. Indeed, throughout the sequence of Figures 3a-c, all isolated fullerenes (unmarked objects) display only spherical symmetry, in contrast to clustered objects (marked by the yellow dashed lines) where internal structure is evident. Upon increasing the electron dose, some $C_{60}$ clearly form bonds with their neighbors, becoming "locked" into a certain position and allowing their structure to be revealed.

Even more remarkable are the few fullerenes that appear to form a bond with a neighbor without fusing together (especially blue and red dashed lines in Figure 3a-b). Consider, for example, the trimer marked with I, II and III in Figure 3a: here, the fullerene I that is bound to two neighbors is presumably locked in its rotation, and shows clear internal structure. However, those having bonded only on one side (II, III) apparently still rotate (or oscillate around a ground state) since their structure is smeared out. Indeed, the appearances of rotating and locked molecules are quite distinct, as revealed by the STEM image simulation shown in Figure 3e; the molecule on the left is an average of 100 random orientations, the one on the right is locked. Based on their appearance, we also conclude that the fullerene dimers or '*peanuts*' (*37*), highlighted by purple dashed lines in Figures 3a-c, exhibit rotation around their longitudinal axis that smears out any internal structure (note the clear difference between the purple marked dimers and polymers with additional cross-linking marked with yellow).

Already one bond between two fullerenes should make the structure rotationally stiff due to the overlap of the *p*-orbitals on the connecting C atoms. However, the energy input from the electron beam can overcome this barrier. Supplementary Video 3 shows a density functional theory molecular dynamics simulation (Materials and Methods) of a dimer where one molecule is prevented from rotating by fixing atoms on its back wall, and one atom on the side of the other molecule is given a high kinetic energy, emulating an electron impact. This mimics the situation where a fullerene is bonded by a single C-C bond to an already rotationally locked neighbor. As can be seen from the video, the simulated impact results in a noticeable rotation of the second molecule even while the bond is preserved.

We further analyzed the dimer bonding by measuring the apparent center-to-center distances ($d$) of paired molecules from Figure 3a. For comparison, Figure 3d shows a pair of free molecules (molecules IV and V), whose center-to-center distance $d'$ is set to 9.6 Å based on the average monolayer spacing. To extract the $d$ of bound molecules, we fitted Gaussian line shapes to the outer edges of the intensity profiles and used the peak-to-peak separation ($D$) to calculate $d = d' + (D - D')$, where $D'$ is the reference distance measured for non-bonded molecules. Figure 3f shows what is possibly a singly bonded $C_{59}$-$C_{59}$ dimer with $d = 9.6$ Å (molecules I and III), closely resembling the simulated equivalent in Figure 3g. (We also simulated the spacing of a singly bonded $C_{60}$-$C_{59}$ and of a $C_{60}$-$C_{60}$ dimer with two bonds, yielding slightly larger values of 9.9 and 10.0 Å, respectively; Supplementary Figure S7). An apparently more tightly bound pair (molecules I and II) is compared in Figures 3h-i: this might be a pair of $C_{59}$ molecules connected by three bonds, based on the measured $d = 8.6$ Å, matching well to a simulated value of 8.7 Å. The peanut structures are compared in Figures 3j-k, where we see that the experimental structure (molecule VI) is slightly smaller than the simulated one, 8.5 Å vs. 8.9 Å. However, stable fullerene peanuts will likely have fewer than 120 atoms and thus many possible structures.

**CONCLUSIONS**

In conclusion, we have created and characterized a mixed-dimensional sandwich heterostructure of $C_{60}$ fullerenes encapsulated by two layers of graphene. The fullerene monolayers inside the sandwich exhibit a lattice spacing of 9.6 Å, ~5% smaller than the bulk spacing. Dynamics of entire molecules can be observed, with weakly bound fullerenes oscillating between different positions at the edges of 2D $C_{60}$ molecular crystals where the graphene layers are suspended over nanometer areas, along with mobile vacancies in disordered 2D fullerene layers. Finally, we observed the transition from rotating individual fullerenes through dimers with suppressed rotation to molecular clusters locked into position, allowing their internal structure to be revealed. The graphene sandwich thus provides a nanoscale reaction chamber, a clean interface to the microscope vacuum, some suppression of radiation damage, and a low contrast background that can be subtracted from the image.

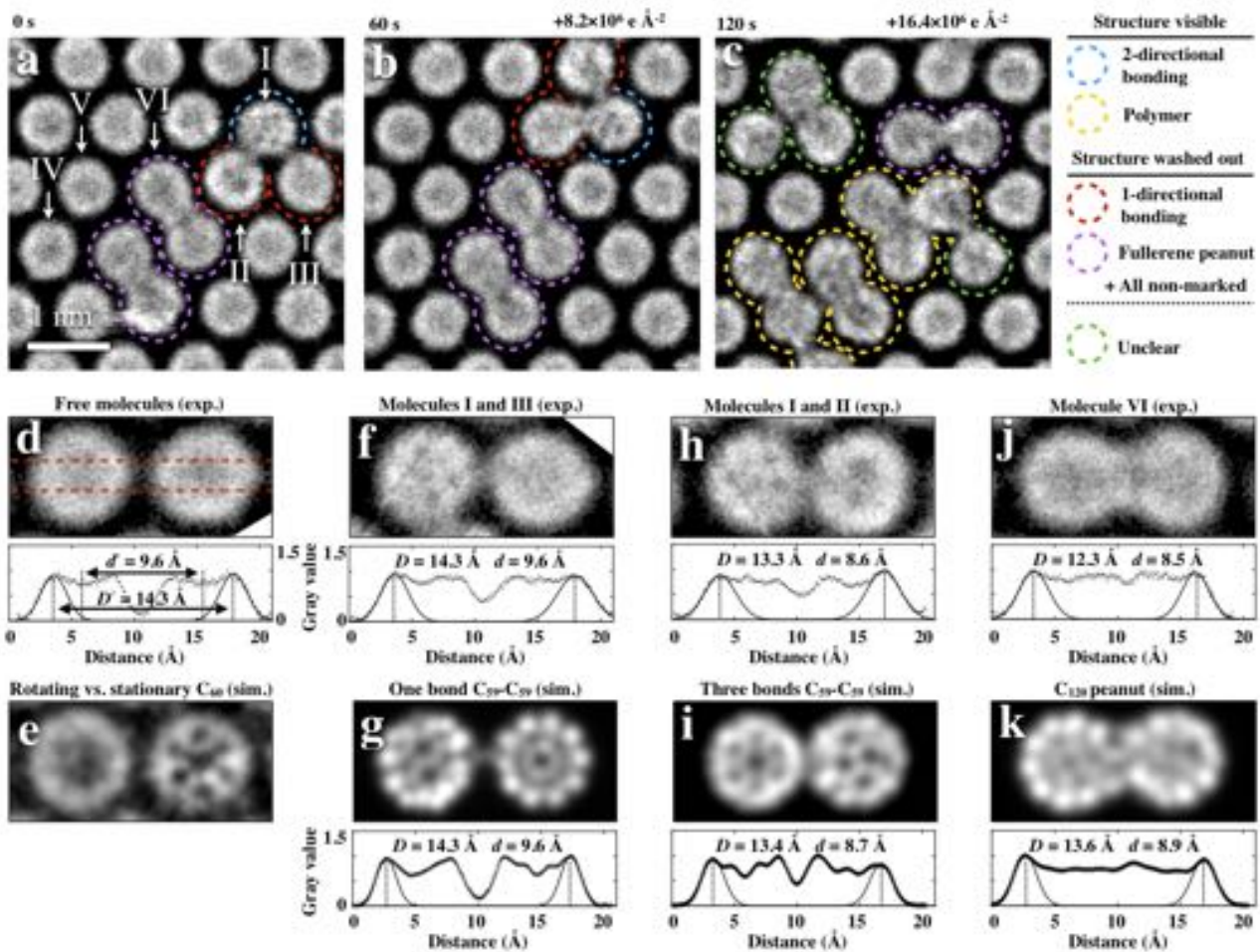

Fig 3. Fullerene reactions. **a-c)** A dose-series of a $C_{60}$ sandwich, smoothed to reduce pixel noise, Fourier-filtered to remove the graphene contribution (Supplementary Figure S4), and contrast enhanced to maximize the visibility of the $C_{60}$ structures (unfiltered images and images with different contrast settings are given as Supplementary Figures S5-S6). The blue dashed lines highlight the $C_{60}$ that appear to be locked in position by two-directional bonding, showing a visible internal structure, and those stationary due to polymerization with yellow color. The dimers (purple) and molecules with a one-directional bond (red) still appear to be moving, obscuring their internal structure. **d)** STEM image of a pair of rotating $C_{60}$ (IV-V in 3a) and a simulation of a rotating (left) and a stationary (right) $C_{60}$ in 3e). **f)** A loosely bound pair of fullerenes and a simulated $C_{59}$-$C_{59}$ dimer with one bond in g), and a tightly bound pair and a simulated $C_{59}$-$C_{59}$ with three bonds in **h-i)**. A typical peanut dimer in j) and an image simulation of $C_{120}$ in k). The red dashes in d) indicate the integration width of the line profiles plotted in panels d-k).

# MATERIALS AND METHODS

## Sandwich fabrication

$C_{60}$ were thermally evaporated at a pressure of ~$10^{-5}$ mbar onto commercial graphene-coated TEM grids from Graphenea Inc. Prior to deposition, the grids were annealed at 400 °C in air for 30 min, and the $C_{60}$ powder for 120 min at 200 °C in vacuum. The evaporation took place in a vacuum chamber (Mantis HEX deposition system) with the target at 100 °C at the rate of ~1 Å/s measured with an ultra-high vacuum compatible quartz-crystal monitor. A second TEM grid used for stacking was placed on top and adhered via the tensile forces resulting from evaporating a droplet of isopropyl alcohol (IPA, semiconductor grade PURANAL, Sigma-Aldrich), followed by drying and then separating the grids with tweezers.

## Scanning transmission electron microscopy (STEM)

Atomic-resolution imaging was conducted using an aberration-corrected Nion UltraSTEM 100 scanning transmission electron microscope operated with 60 keV primary beam energy, with near-ultrahigh vacuum of $2\times10^{-7}$ Pa at the sample. The angular range for the medium angle annular dark-field (MAADF) detector was 60–200 mrad.

## Raman spectroscopy

Raman spectra (Supplementary Figure S1e) were acquired using an ND-MDT NTEGRA Spectra AFM/Raman instrument with a 473 nm excitation wavelength (2.62 eV) under ambient conditions. The output laser power was ~4 mW over a focused spot with a diameter of ~0.5 µm.

## STEM simulations

STEM simulations used QSTEM version 2.31 with a chromatic aberration coefficient of 1 mm, a spherical aberration coefficient of 1 µm, energy spread of 0.48 eV, and MAADF detector angle range set to the experimental range of 60–200 mrad. The illumination semi-angle was 35 mrad. In Figure 3e, to mimic the experimental contrast of the rotating fullerene, graphene was included in the simulations and later Fourier filtered.

## Atomistic simulations

Most simulations were carried out with the LAMMPS code (*38*), using an AIREBO-potential augmented by a Morse potential for the intermolecular interaction (A-M) to describe the covalent and vdW interactions, respectively (*39, 40*). To study the intermolecular spacing and diffusion barriers in sandwiched $C_{60}$ layers, a model supercell consisting of seven rows of $C_{60}$ with periodic boundary conditions along the over- and underlying graphene sheet edges. The misorientation angle between the graphene lattices was set to 11.6°, resulting in a moiré pattern periodicity of ~12 Å matching the experiments. This unit cell was repeated by 5 and 20 times, resulting in a supercell consisting of 33820 carbon atoms, including the $C_{60}$ molecules. The total energy was minimized by relaxing the layers without constraints until the forces were < $10^{-4}$ eV Å$^{-1}$.

For bulk $C_{60}$ crystals, we additionally compared these results with density functional theory (DFT), calculated using two different van der Waals functionals (C09 and DF2) (*41, 42*). We relaxed the minimal single-molecule unit cells of both 3D FCC and 2D crystallites using the planewave mode in the GPAW simulation package (*43*) (600 eV cutoff energy, and 5x5x1 and 5x5x5 Monkhorst-Pack k-point meshes for the 2D and 3D cases, respectively). The results agreed with the A-M calculation in that there is no significant difference between the 2D and 3D spacing of the fullerenes in absence of the graphene sandwich. For 3D bulk $C_{60}$, we found zero Kelvin equilibrium distances of 9.65 Å (A-M), 9.86 Å (C09) and 10.03 Å (DF2), whereas for an infinite 2D sheet, the distances are 9.62 Å (A-M, consistent with ref. (*44*)), 9.91 Å (C09) and 10.02 Å (DF2).

For the diffusion barrier simulations, we cut out part of the fullerene-containing area of the model, created a gap inside the fullerene monolayer, and fixed the outermost rows of fullerenes and the edges of the graphene layers. The barriers were calculated by using the nudged elastic band method(*45*) with a spring constant of 10 eV/Å.

To study the bonding of the fullerenes, we placed two $C_{60}$ in the center of a 32x20x20 Å unit cell and relaxed their atomic structure using the C09 functional in the GPAW FD mode (0.19 Å grid spacing, Gamma point only) (*46*). To study the effect of irradiation on apparent bond length, we removed first one C atom from the second $C_{60}$ and then relaxed the structure, and then further removed another C atom from the first $C_{60}$ and relaxed that structure, resulting in dimers either with one or three bonds. For the $C_{60}$-$C_{59}$ system, we further ran molecular dynamics simulations in the LCAO DFT mode (15 eV kinetic energy, Supplementary Video 3) to simulate the rotation of the $C_{59}$ caused by an electron impact on one of the carbon atoms.

## SUPPLEMENTARY MATERIAL

Supplementary material for this article is available at

Fig. S1. Overview scanning transmission electron microscope (STEM) micrographs, electron diffraction patterns and Raman spectra.

Fig. S2. Structure of $C_{60}$ multilayers.

Fig. S3. Moiré contrast at the edges of the $C_{60}$ monolayer.

Fig. S4. An example of Fourier filtering.

Fig. S5. Original STEM micrographs of the manuscript Figures 3a-c.

Fig. S6. Fourier filtered views of the rotating and fusing fullerenes from Figure 3a-c without any overlay and with varying contrast settings.

Fig. S7. STEM simulations of a singly bonded $C_{60}$-$C_{59}$ and of a $C_{60}$-$C_{60}$ with two bonds.

Movie S1. $C_{60}$ diffusing away from a row of molecules enveloped by graphene.

Movie S2. $C_{60}$ diffusing along a row of molecules enveloped by graphene.

Movie S3. Bond rotation in fullerene dimer after a 15 eV electron impact.

References (*16, 29, 47-49*)

## REFERENCES AND NOTES

1. K. S. Novoselov *et al.*, Electric Field Effect in Atomically Thin Carbon Films. *Science* **306**, 666-669 (2004).
2. A. K. Geim, I. V. Grigorieva, Van der Waals heterostructures. *Nature* **499**, 419-425 (2013).
3. D. Jariwala, T. J. Marks, M. C. Hersam, Mixed-dimensional van der Waals heterostructures. *Nat Mater* **16**, 170-181 (2017).
4. B. W. Smith, M. Monthioux, D. E. Luzzi, Encapsulated $C_{60}$ in carbon nanotubes. *Nature* **396**, 323-324 (1998).
5. H. W. Kroto, J. R. Heath, S. C. O'Brien, R. F. Curl, R. E. Smalley, $C_{60}$: Buckminsterfullerene. *Nature* **318**, 162-163 (1985).
6. Q. Wang, R. Kitaura, Y. Yamamoto, S. Arai, H. Shinohara, Synthesis and TEM structural characterization of $C_{60}$-flattened carbon nanotube nanopeapods. *Nano Res*. **7**, 1843-1848 (2014).
7. H. R. Barzegar *et al.*, $C_{60}$/Collapsed Carbon Nanotube Hybrids: A Variant of Peapods. *Nano Letters* **15**, 829-834 (2015).
8. T. Rabenau, A. Simon, R. K. Kremer, E. Sohmen, The energy gaps of fullerene $C_{60}$ and $C_{70}$ determined from the temperature dependent microwave conductivity. *Zeitschrift für Physik B Condensed Matter* **90**, 69-72 (1993).
9. C. T. Chen *et al.*, Electronic states and phases of $K_xC_{60}$ from photoemission and X-ray absorption spectroscopy. *Nature* **352**, 603-605 (1991).
10. L. W. Tutt, A. Kost, Optical limiting performance of $C_{60}$ and $C_{70}$ solutions. *Nature* **356**, 225-226 (1992).
11. P.-M. Allemand *et al.*, Molecular Soft Ferromagnetism in a Fullerene $C_{60}$. *Science* **253**, 301-302 (1991).

12. F. Wudl, J. D. Thompson, Buckminsterfullerene $C_{60}$ and organic ferromagnetism. *Journal of Physics and Chemistry of Solids* **53**, 1449-1455 (1992).
13. A. Chuvilin, U. Kaiser, E. Bichoutskaia, N. A. Besley, A. N. Khlobystov, Direct transformation of graphene to fullerene. *Nat Chem* **2**, 450-453 (2010).
14. M. Koshino *et al.*, Analysis of the reactivity and selectivity of fullerene dimerization reactions at the atomic level. *Nat Chem* **2**, 117-124 (2010).
15. M. Terrones, Transmission electron microscopy: Visualizing fullerene chemistry. *Nat Chem* **2**, 82-83 (2010).
16. K. Kim *et al.*, Structural and Electrical Investigation of $C_{60}$–Graphene Vertical Heterostructures. *ACS Nano* **9**, 5922-5928 (2015).
17. G. Li *et al.*, Self-assembly of $C_{60}$ monolayer on epitaxially grown, nanostructured graphene on Ru(0001) surface. *Applied Physics Letters* **100**, 013304 (2012).
18. P. Utko, J. Nygård, M. Monthioux, L. Noé, Sub-Kelvin transport spectroscopy of fullerene peapod quantum dots. *Applied Physics Letters* **89**, 233118 (2006).
19. S. Sato, T. Yamasaki, H. Isobe, Solid-state structures of peapod bearings composed of finite single-wall carbon nanotube and fullerene molecules. *Proceedings of the National Academy of Sciences* **111**, 8374-8379 (2014).
20. S. Saito, A. Oshiyama, Design of $C_{60}$-graphite cointercalation compounds. *Physical Review B* **49**, 17413-17419 (1994).
21. M. S. Fuhrer, J. G. Hou, X. D. Xiang, A. Zettl, C60 intercalated graphite: Predictions and experiments. *Solid State Communications* **90**, 357-360 (1994).
22. A. F. Hebard *et al.*, Superconductivity at 18 K in potassium-doped $C_{60}$. *Nature* **350**, 600-601 (1991).
23. A. R. Kortan *et al.*, Superconductivity at 8.4 K in calcium-doped $C_{60}$. *Nature* **355**, 529-532 (1992).
24. X. Ye *et al.*, Single-particle mapping of nonequilibrium nanocrystal transformations. *Science* **354**, 874-877 (2016).
25. J. M. Yuk *et al.*, High-Resolution EM of Colloidal Nanocrystal Growth Using Graphene Liquid Cells. *Science* **336**, 61-64 (2012).
26. G. Algara-Siller, S. Kurasch, M. Sedighi, O. Lehtinen, U. Kaiser, The pristine atomic structure of $MoS_2$ monolayer protected from electron radiation damage by graphene. *Applied Physics Letters* **103**, 203107 (2013).
27. R. Zan *et al.*, Control of Radiation Damage in $MoS_2$ by Graphene Encapsulation. *ACS Nano* **7**, 10167-10174 (2013).
28. K. Kurotobi, Y. Murata, A Single Molecule of Water Encapsulated in Fullerene $C_{60}$. *Science* **333**, 613-616 (2011).
29. G. Van Tendeloo *et al.*, Phase transitions in fullerene ($C_{60}$) and the related microstructure: a study by electron diffraction and electron microscopy. *The Journal of Physical Chemistry* **96**, 7424-7430 (1992).
30. G. Zhang, R. Zhou, X. C. Zeng, Carbon nanotube and boron nitride nanotube hosted $C_{60}$-V nanopeapods. *Journal of Materials Chemistry C* **1**, 4518-4526 (2013).
31. K. Hirahara *et al.*, One-Dimensional Metallofullerene Crystal Generated Inside Single-Walled Carbon Nanotubes. *Physical Review Letters* **85**, 5384-5387 (2000).
32. S. Bommel *et al.*, Unravelling the multilayer growth of the fullerene C60 in real time. *Nature Communications* **5**, 5388 (2014).
33. H. Liu, Z. Lin, L. V. Zhigilei, P. Reinke, Fractal Structures in Fullerene Layers: Simulation of the Growth Process. *The Journal of Physical Chemistry C* **112**, 4687-4695 (2008).
34. A. V. Talyzin *et al.*, Synthesis of $C_{59}H_x$ and $C_{58}H_x$ Fullerenes Stabilized by Hydrogen. *The Journal of Physical Chemistry B* **109**, 5403-5405 (2005).

35. T. Susi *et al.*, Isotope analysis in the transmission electron microscope. *Nature Communications* **7**, 13040 (2016).
36. M. Neek-Amal, N. Abedpour, S. N. Rasuli, A. Naji, M. R. Ejtehadi, Diffusive motion of $C_{60}$ on a graphene sheet. *Physical Review E* **82**, 051605 (2010).
37. J. Onoe, T. Nakayama, M. Aono, T. Hara, Structural and electrical properties of an electron-beam-irradiated $C_{60}$ film. *Applied Physics Letters* **82**, 595-597 (2003).
38. S. Plimpton, Fast Parallel Algorithms for Short-Range Molecular Dynamics. *Journal of Computational Physics* **117**, 1-19 (1995).
39. T. C. O'Connor, J. Andzelm, M. O. Robbins, AIREBO-M: A reactive model for hydrocarbons at extreme pressures. *The Journal of Chemical Physics* **142**, 024903 (2015).
40. D. W. Brenner, Empirical potential for hydrocarbons for use in simulating the chemical vapor deposition of diamond films. *Physical Review B* **42**, 9458-9471 (1990).
41. V. R. Cooper, Van der Waals density functional: An appropriate exchange functional. *Phys. Rev. B* **81**, 161104 (2010).
42. K. Lee, E. D. Murray, L. Kong, B. I. Lundqvist, D. C. Langreth, Higher-accuracy van der Waals density functional. *Phys. Rev. B* **82**, 081101 (2010).
43. J. Enkovaara *et al.*, Electronic structure calculations with GPAW: a real-space implementation of the projector augmented-wave method. *Journal of Physics: Condensed Matter* **22**, 253202 (2010).
44. C. D. Reddy, Z. Gen Yu, Y.-W. Zhang, Two-dimensional van der Waals $C_{60}$ molecular crystal. *Scientific Reports* **5**, 12221 (2015).
45. G. Henkelman, B. P. Uberuaga, H. Jónsson, A climbing image nudged elastic band method for finding saddle points and minimum energy paths. *The Journal of Chemical Physics* **113**, 9901-9904 (2000).
46. A. H. Larsen, M. Vanin, J. J. Mortensen, K. S. Thygesen, K. W. Jacobsen, Localized atomic basis set in the projector augmented wave method. *Physical Review B* **80**, 195112 (2009).
47. S. Liu, Y.-J. Lu, M. M. Kappes, J. A. Ibers, The Structure of the $C_{60}$ Molecule: X-Ray Crystal Structure Determination of a Twin at 110 K. *Science* **254**, 408-410 (1991).
48. M. S. Dresselhaus, G. Dresselhaus, P. C. Eklund, Raman Scattering in Fullerenes. *Journal of Raman Spectroscopy* **27**, 351-371 (1996).
49. L. M. Malard, M. A. Pimenta, G. Dresselhaus, M. S. Dresselhaus, Raman spectroscopy in graphene. *Physics Reports* **473**, 51-87 (2009).

**Funding**

This work was supported by the European Research Council Starting Grant No. 336453-PICOMAT and Austrian Science Fund (FWF) under Project Numbers P 25721-N20, I1283-N20 and P 28322-N36. K.M. acknowledges the support from the Finnish Foundations' Post Doc Pool and T.J.P. funding from the European Union's Horizon 2020 research and innovation programme under the Marie Skłodowska-Curie grant agreement No. 655760—DIGIPHASE. J.K. acknowledges funding from Wiener Wissenschafts- Forschungs- und Technologiefonds through project MA14-009.

**Acknowledgements**

Computational resources from the Vienna Scientific Cluster are gratefully acknowledged.

**Authors contributions**

R.M. and A.M. prepared samples. R.M., A.M., T.J.P., and C.M. performed STEM experiments. R.M. performed Raman spectroscopy. K.M., R.M. and J.C.M. analyzed experimental data. M.R., A.M., T.S., J.K. and K.M. carried out simulations and analyzed results. K.M., T.S., J.K., R.M. and J.C.M. wrote the paper with input and comments from all authors. J.C.M. conceived and supervised the study with co-supervision from J.K and T.S.

**Competing interests:** The authors declare that they have no competing interests.

**Data and materials availability:** All data needed to evaluate the conclusions in the paper are presented in the paper and/or the Supplementary Materials. Additional data related to this paper may be requested from the authors.

# Supplementary Material for

**Buckyball Sandwiches**

R. Mirzayev*, K. Mustonen* [a)], M.R.A. Monazam, A. Mittelberger, T.J. Pennycook, C. Mangler, T. Susi, J. Kotakoski and J.C. Meyer [b)]

*University of Vienna, Faculty of Physics, Boltzmanngasse 5, A-1090 Vienna, Austria*

___________________________

Electronic mail: [a)] kimmo.mustonen@univie.ac.at, [b)] jannik.meyer@univie.ac.at

* These authors contributed equally to this work

# SAMPLE MORPHOLOGY

Figure S1a shows an overview of $C_{60}$ crystallites on a single layer of graphene, and Figure 1b shows $C_{60}$ crystallites in a graphene sandwich. In both cases, the fullerenes form islands from a few tens up to one hundred nanometers in diameter, with thicknesses up to ~10 layers (visible as the different contrast of each island, especially in Figure S1a).

The first confirmation of the sandwich structure is obtained through electron diffraction patterns shown in Figures S1c and 1d. The corner points of the superimposed hexagons correspond to the alignment of the $C_{60}$ crystallites and the graphene lattice, revealing the presence of two misaligned graphene layers in Figure S1d. The two constituent lattices—graphene and $C_{60}$ crystallite—share the same threefold symmetry as depicted in Figures S1f and 1g. We can therefore calculate the $C_{60}$ lattice spacing relative to the well-known graphene lattice constant (2.46 Å), yielding 10.4±0.3 Å for both non-sandwiched and sandwiched multi-layer $C_{60}$ structures. This is consistent with the accepted lattice constants of bulk crystallites from electron and x-ray diffraction techniques *(14, 25)*.

Raman spectroscopy provides additional evidence of the heterostructure. In Figure S1e we compare the prominent Raman features of sandwiched and non-sandwiched $C_{60}$. These are the pentagon pinch mode at ~1469 cm$^{-1}$ (the atoms in carbon pentagons tangentially oscillating in phase) and two additional modes, the squashing at ~273 cm$^{-1}$ and breathing at ~497 cm$^{-1}$, related to radial expansions and contractions. Interestingly, all of these modes were suppressed by the added sandwiching layer: pentagon pinch by ~72% and squashing by ~78%, while the breathing mode *(42)* —usually more pronounced than squashing in suspensions — dropped below the noise floor.

The Raman modes of graphene are also affected. In Figure S1e, the first order hexagonal vibration is marked as G at ~1580 cm$^{-1}$, the second order as 2D at ~2690 cm$^{-1}$ and the disorder mode at ~1360 cm$^{-1}$ as D. The sandwiching altered the relative intensity of the G and 2D modes, with the latter likely being dispersed due to the interlayer coupling resulting in the full width at half maximum (FWHM) increasing from 44 to 50 cm$^{-1}$ *(43)*. The graphene disorder-related D mode was detected exclusively in sandwiches, although its origin remains unclear.

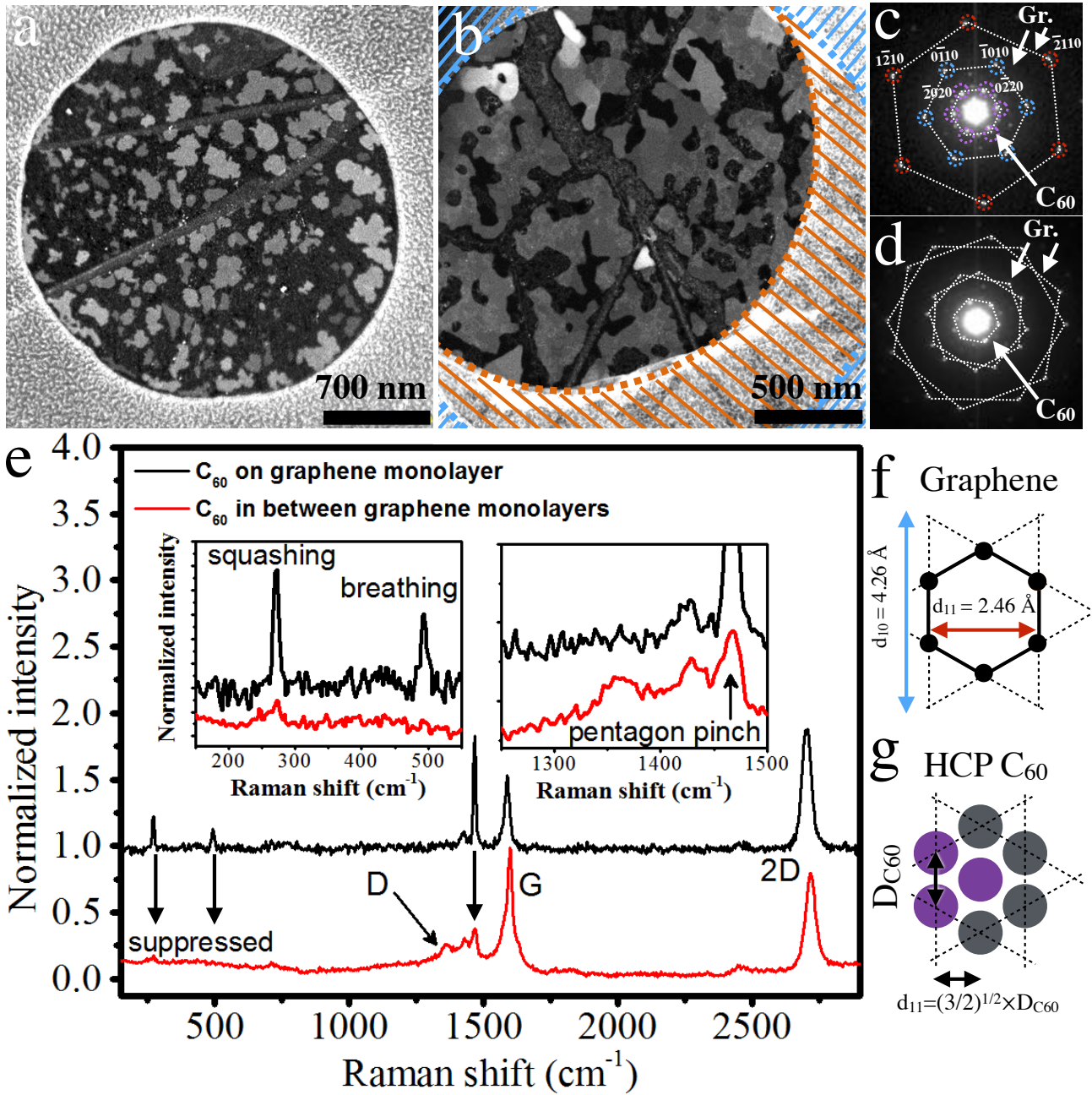

Fig. S1. Overview STEM micrographs, electron diffraction patterns and Raman spectra. **a)** STEM micrograph of a graphene monolayer covered with $C_{60}$ and **b)** of a buckyball sandwich. The overlapping holes in the TEM grid support films are highlighted with overlaid orange and blue hatching. **c)** Electron diffraction pattern of graphene covered with $C_{60}$ and **d)** of a sandwich structure, representing an average signal from an area of ca. $7 \times 10^4$ nm$^2$. The colored dashed circles denote the Miller-Bravais indices of the diffraction components of graphene (Gr.; red and blue) and the $C_{60}$ (purple). **e)** Raman spectra of $C_{60}$ on graphene and in a sandwich. The insets show the $C_{60}$ modes in greater detail. **f-g)** The triangular symmetry of graphene (f) and of $C_{60}$ crystallites (g). The dashed lines indicate the (111) Bravais planes, and the coloring corresponds to (c-d).

# C$_{60}$ MULTILAYERS

High-resolution STEM imaging reveals continuous and highly ordered layers of C$_{60}$ on both single-layer graphene and in sandwiches. Figure S2a shows a representative 34×34 nm$^2$ field of view from non-sandwiched C$_{60}$ deposited on a single layer of graphene, and Figure S2b a higher magnification from the region marked as [b]. The STEM micrograph representative of the sandwiched multilayers shown in Figure S2e shows much clearer contrast, allowing the two distinct stacking orders to be distinguished. We could find perfectly clean regions for imaging the fullerenes only in the sandwich structures. The uncovered fullerene surfaces always had some amount of amorphous contamination. Based on the STEM image contrast and by using fullerene mono-layers and graphene as reference, the structure in Fig. S2e can be identified as a fullerene triple-layer.

Although not obvious to the unaided eye, Figures S2a and S2e also contain traces of the underlying graphene. The lattice can be extracted by calculating the Fourier transform (FT) of the images, shown as Figures S2c and S2f (to guide the eye, dashed circles are used to mark the positions of the discernible graphene diffraction components, and solid hexagons are used to indicate the orientations of C$_{60}$ and graphene (111) planes. The graphene lattice in the FT can then be used as an accurate reference that allows us to measure the fullerene spacing in both cases to be 10.0±0.3 Å, consistent with the value obtained above from the diffraction patterns. The mismatch angle between the C$_{60}$ and graphene lattices is also interesting. As evidenced by the histogram shown in Figure S2d, the C$_{60}$ crystallites seem to organize epitaxially on graphene, preferring either the graphene zig-zag (ZZ, 0°) or armchair (AC, 30°) directions. This tendency has been earlier explained via *ab initio* calculations of vdW-favored adsorption geometries *(44)*. However, when the sandwiching layer is added at a random angle, the second graphene layer can no longer find a favored registry with either the first layer or the fullerenes. This is illustrated by the stacked histogram of Figure S2g, where we plot the smaller ($\Theta_{Gr-C60-I}$) and the larger ($\Theta_{Gr-C60-II}$) mismatch angle for the sandwiches. The distribution of the smaller angle is close to what is observed for a single graphene layer, whereas the larger angle has a more uniform distribution consistent with random variation.

Finally, to confirm the identity of the projected structure, we constructed C$_{60}$ trilayers for STEM image simulations (Figure S2h). According to these simulations and earlier work, an ABC *(44)* stacked multilayer results in a honeycomb-like periodicity similar to the one observed in the upper left corner of Figure S2e. The simulation and the area marked with [i] are compared in Figure S2i, clearly identifying the ABC stacking. The lower right part of Figure S2e exhibits a different contrast, with Figure S2j showing a cutout of the area marked with [j]. The ABA trilayer simulation next to it again perfectly matches the experimental contrast. Similarly, it can be seen that the $C_{60}$ in the non-sandwiched multilayer of Figure S2b are arranged in an ABA lattice.

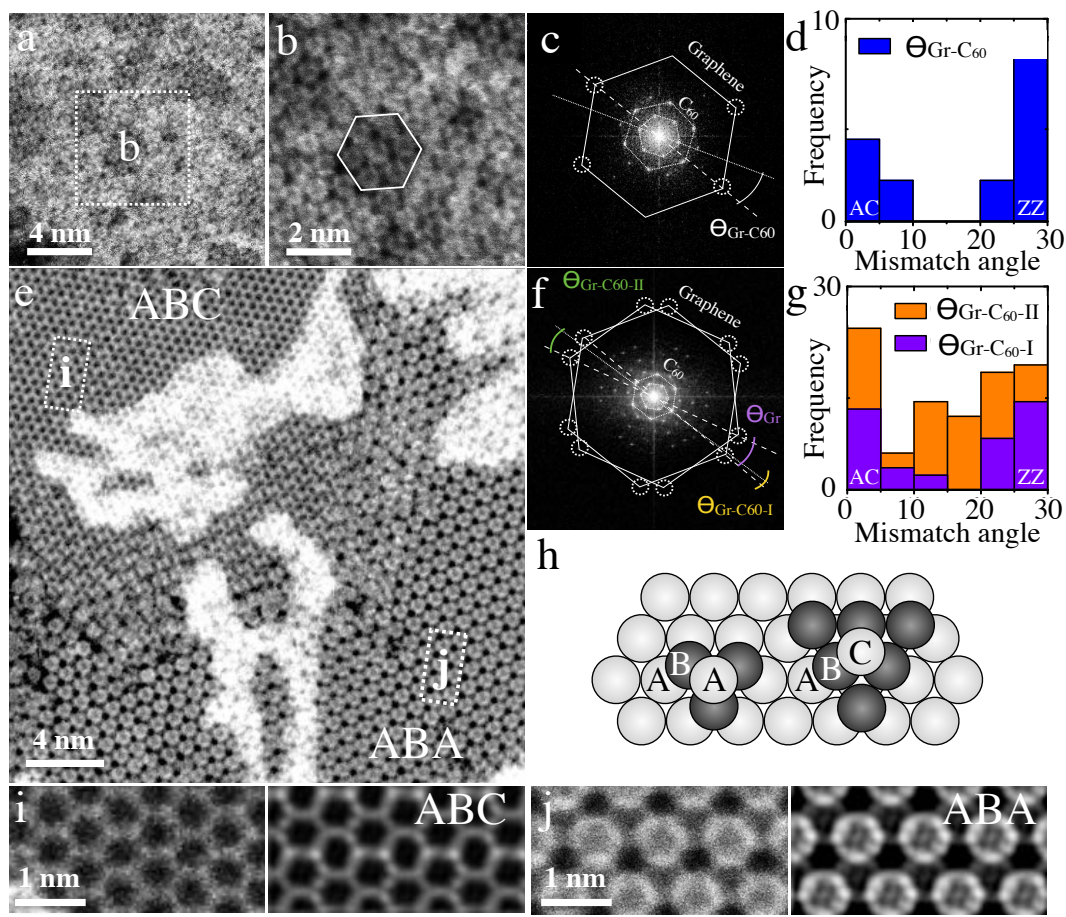

Fig. S2. Structure of $C_{60}$ multilayers: **a-d)** non-sandwiched multilayer, **e-g)** sandwiched multilayer. **a)** STEM micrograph of a multilayer area with overlying carbon contamination. The relatively clean area marked by [b] shown in **b)** still exhibits a disordered amorphous coverage. **c)** A Fourier transform of b) showing the orientation of graphene and $C_{60}$ (111) planes in reciprocal space. **d)** Histogram plotting the frequency of relative orientation of $C_{60}$ and graphene (111) planes (statistics gathered from diffraction patterns). **e)** A buckyball sandwich with multilayered regions exhibiting two distinct stacking orders. **f)** A Fourier transform of e) showing the orientations of the two graphene layers and $C_{60}$. **g)** Stacked histogram of the mismatch angle distribution between the graphene and fullerene layers. **h)** A schematic illustration of the ABA and ABC stacking orders. **i)** Closeup of the area marked by [i] in panel e) showing a honeycomb pattern that precisely matches the STEM image simulation of ABC stacking shown on the right. **j)** Closeup of the area marked by [j] in panel e), precisely matching the simulated ABA stacking contrast.

# STEM CONTRAST OF THE SANDWICH

STEM images of the edges of a fullerene monolayer (Fig. S3a) provide an additional confirmation that the fullerenes are captured in-between the graphene sheets (rather than e.g. on top of the graphene double layer). In regions devoid of fullerenes, the two graphene layers form a rotated bilayer and display the characteristic moiré pattern in the medium angle annular dark field (MAADF) images. As can be seen in Fig. S3a the pattern is composed of periodic spots of enhanced intensity such as those highlighted by dashed circles in the figure. The STEM simulations of Figures S3b-d show that this contrast is due to a nonlinear effect enhancing the intensity when two atoms are at the same projected position and separated by no more than ca. 5 Å in the beam direction. This pattern is visible over all bilayer areas but changes significantly near the edge of the fullerene layer. Near the edges of the fullerene layers, the sheets separate, causing the moiré pattern to disappear. An experimental image is shown in Fig. S3a, and a comparison to a STEM simulation for a relaxed structure (as in Fig. 2 of the manuscript) is shown in Fig. S3b. While the qualitative agreement is good, the distance between the fullerene edge and the moiré of the rotated graphene bi-layer is slightly larger in the experiment. This can be resolved by adding a strain of 330GPa in the simulation (Fig. 3c), by considering an uneven edge (Fig. 3d shows a relaxed structure with three missing molecules), or a combination of both.

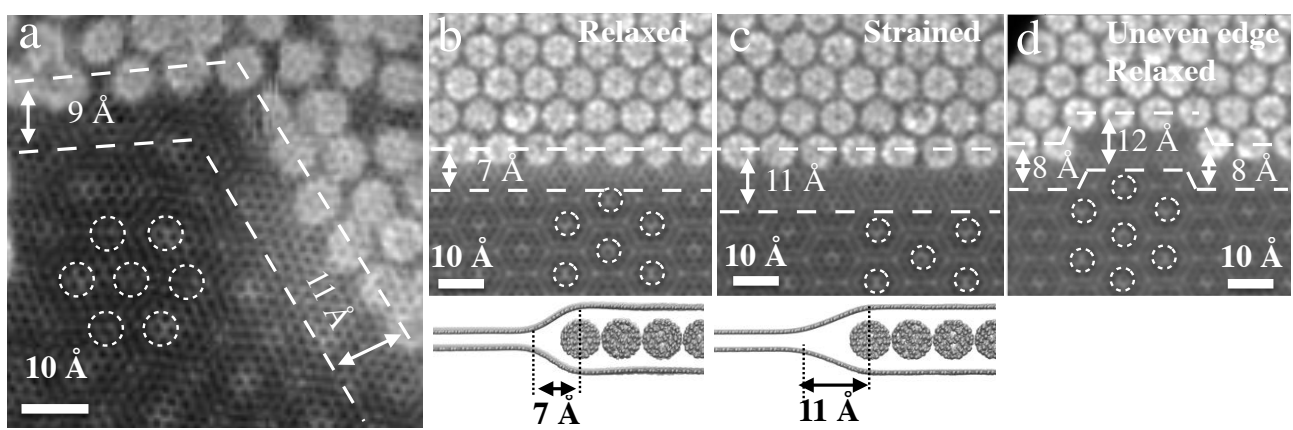

Fig. S3. Moiré contrast at the edges of the $C_{60}$ monolayer. **a)** Experimental image, with the non-linear contribution of the moiré (which indicates a rotated bi-layer graphene) highlighted by dashed circles and the distance of this effect from the fullerene layer given between straight dashed lines. **b)** Simulated STEM image for the relaxed structure, showing slightly smaller distance than the experiment. **c)** Strained structure (330 MPa in vertical direction), and **d)** relaxed structure with an uneven edge, both of which provide a good match to the experiment.

## FOURIER FILTERING

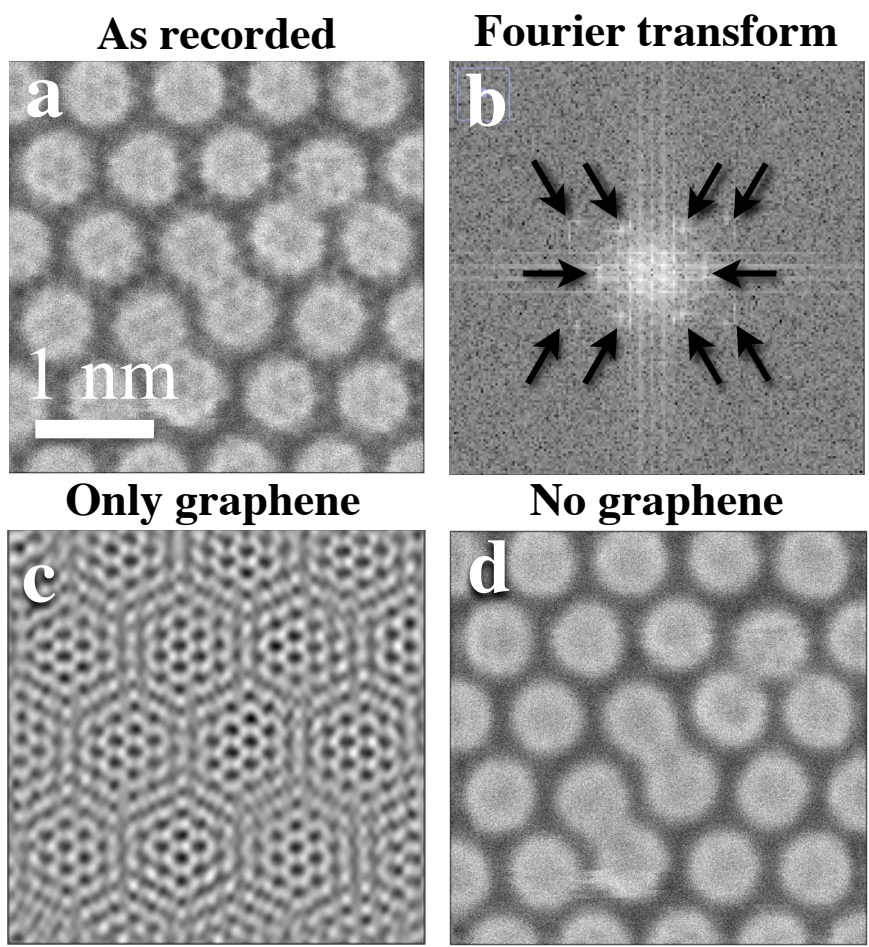

Fig. S4. Fourier filtering. **a)** As-recorded STEM micrograph of a $C_{60}$ sandwich and **b)** its Fourier transform (FT). The arrows identify the frequency components of graphene, which are removed from the filtered image. **c)** The graphene periodic components and **d)** the difference of a) and c).

## UNFILTERED STEM IMAGES OF THE $C_{60}$ SANDWICH

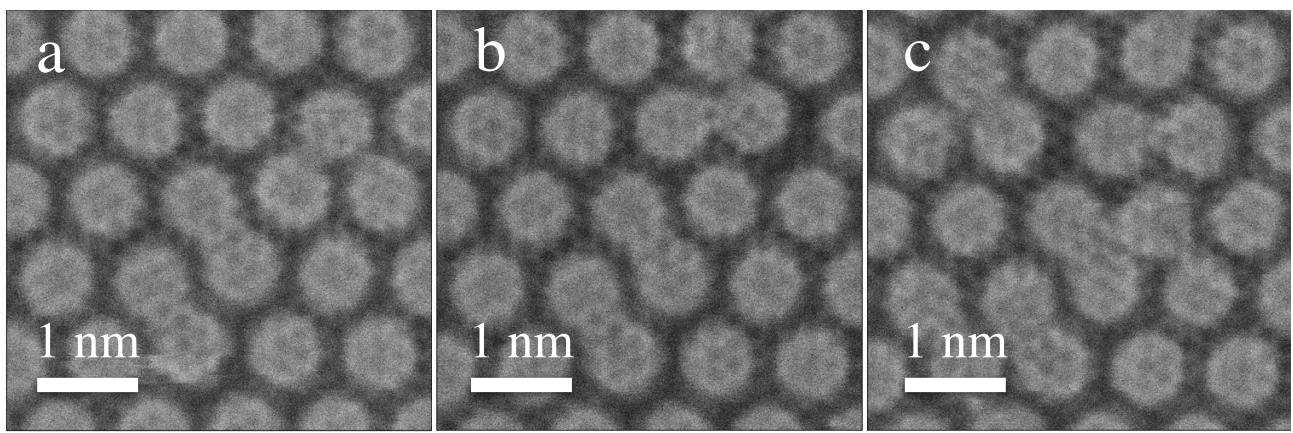

Fig. S5. **a-c)** The as acquired STEM micrographs of the sandwiched monolayer.

# ADDITIONAL FILTERED VIEWS OF THE $C_{60}$ SANDWICH

To demonstrate clearly that some of the fullerenes (mostly the merged ones) display internal structures while others (especially isolated ones) do not, we show here again the Fourier-filtered images (as in Figure 3 of the main manuscript) with three different contrast settings (the best visibility may depend on the settings of the printer or monitor).

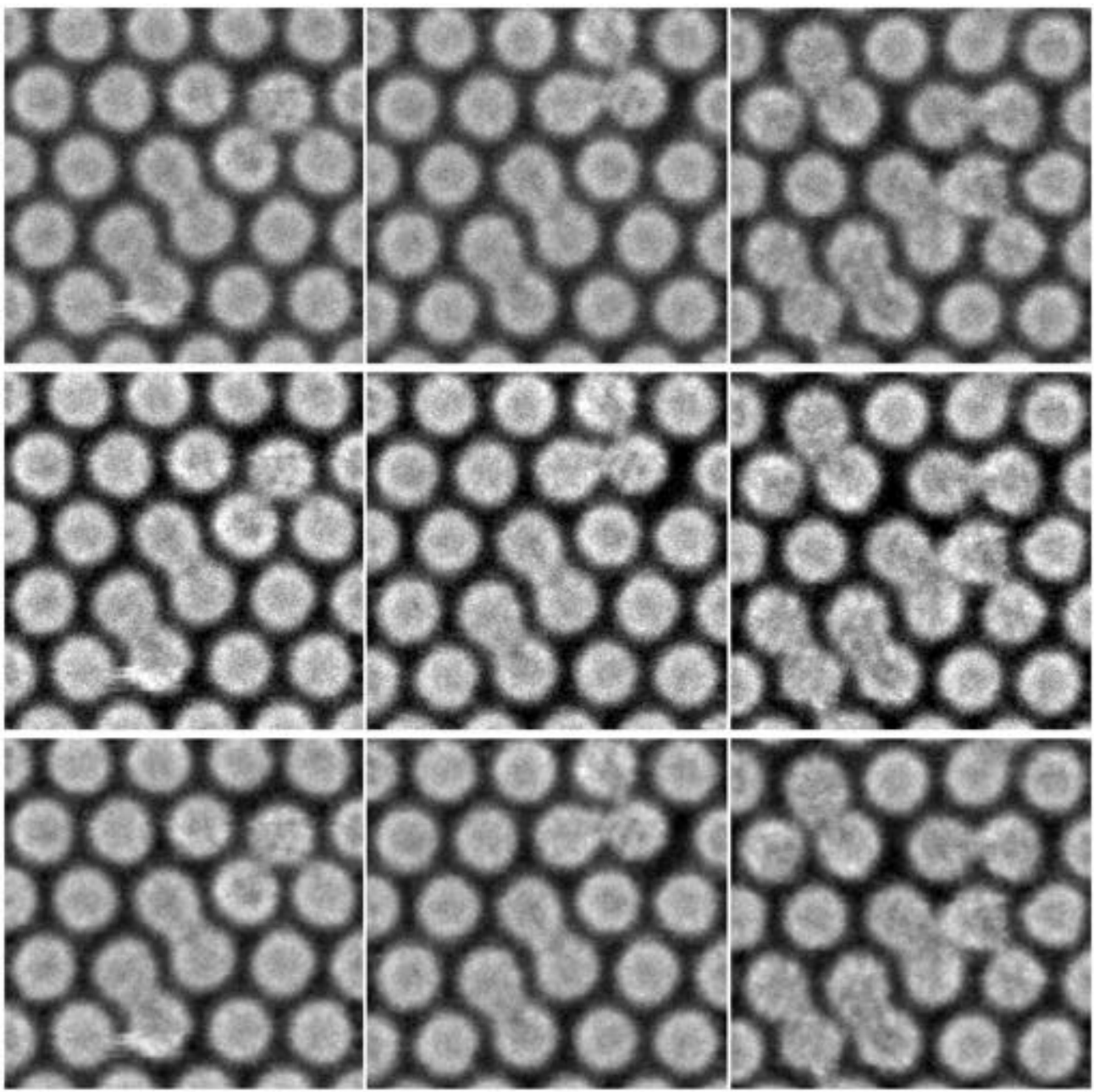

Fig. S6. Fourier filtered views of the rotating and fusing fullerenes. These are the same images as in manuscript Figures 3e-g, only displayed without any overlay and in each line with different contrast settings.

**STEM SIMULATIONS OF A C$_{60}$-C$_{60}$ WITH TWO BONDS AND SINGLY BONDED C$_{60}$-C$_{59}$**

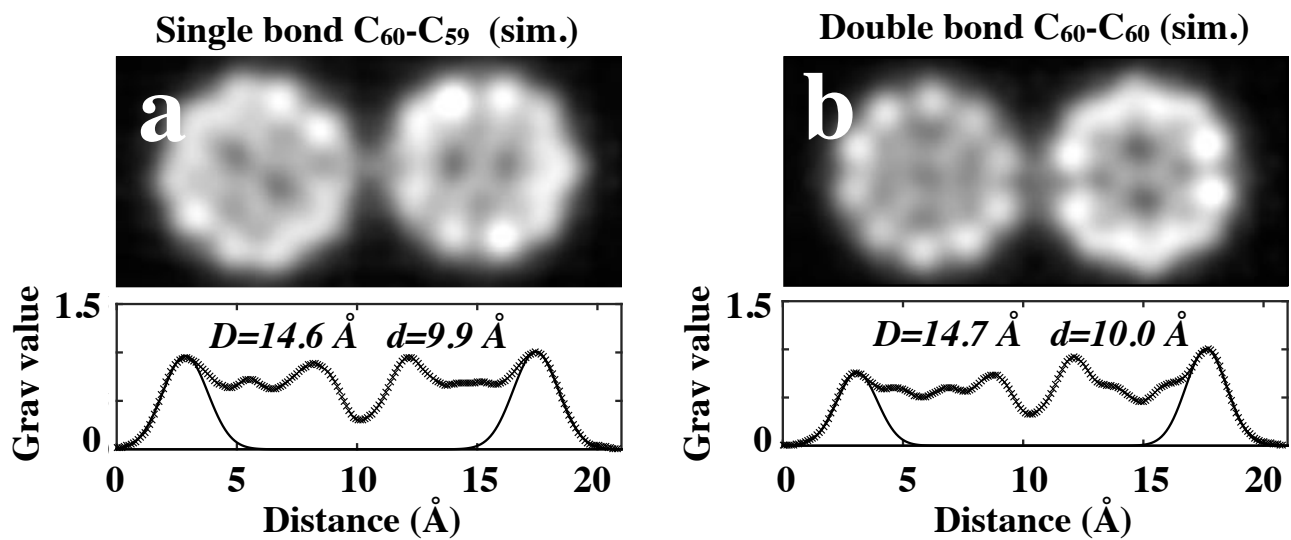

Fig. S7. **a)** STEM simulation of a singly bonded C$_{60}$-C$_{59}$ and **b)** of a C$_{60}$-C$_{60}$ with two bonds.

9